# 2.4GHZ CLASS AB POWER AMPLIFIER FOR HEALTHCARE APPLICATION

Wei Cai[1], Liang Huang[2] and WuJie Wen[3]

[1]Department of Electrical Engineering and Computer Science, University of California, Irvine, CA, USA
*caiw2@uci.edu*

[2]Department of Information & Electronic Engineering, ZheJiang Gongshang University, Hang Zhou, Zhejiang, China
*huangliang@zjgsu.edu.cn*

[3]Department of Electrical Engineering and Computer Science, Florida International University, Miami, FL, USA
*wwen@fiu.edu*

***ABSTRACT***

*The objective of this research was to design a 2.4 GHz class AB Power Amplifier (PA), with 0.18um Semiconductor Manufacturing International Corporation (SMIC) CMOS technology by using Cadence software, for health care applications. The ultimate goal for such application is to minimize the trade-offs between performance and cost, and between performance and low power consumption design. This paper introduces the design of a 2.4GHz class AB power amplifier which consists of two stage amplifiers. This power amplifier can transmit 10dBm output power to a 50Ω load. The power added efficiency is 7.5% at 1dB compression point and the power gain is 10dB, the total power consumption is 0.135W. The performance of the power amplifier meets the specification requirements of the desired.*

***KEYWORDS***

*Two stage, Class AB, Power amplifier, Healthcare*

## 1.INTRODUCTION

Wireless medical sensor networks have offered significant improvements to the healthcare industry in the 21st century. Devices are arranged on a patient's body and can be used to closely monitor the physiological condition of patients. These medical sensors monitor the patient's vital body parameters, such as temperature, heart rate, blood pressure, oxygen saturation, and transmit the data to a doctor in real time [1]. When a doctor reviews the transmitted sensor readings, they can get a better understanding of a patient's health conditions. The benefit for the patients is that they do not need to frequently visit the hospital, thus patients could reap time and money savings. Such wireless medical sensors will continue to play a central role in the future of modern healthcare. People living in rural areas would especially benefit, since 9% of physicians work in rural areas while almost 20% of the US population lives there [2]. A shortage of physicians and specialist is a big issue in such areas, even today. But Wireless Medical Sensor Network technology has the potential to alleviate the problem.

In a wireless sensor network, as seen in the figure 1 below, each device is capable of monitoring, sensing, and/or displaying information. A sensor node is capable of gathering sensory information, processing it in some manner, and communicating with other nodes in the network.

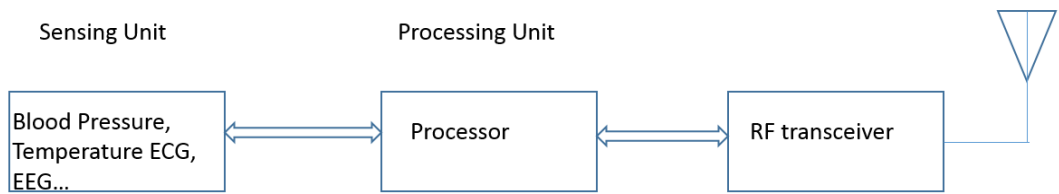

Figure 1. Block diagram of a typical sensor node

Figure 1 shows that the basic sensing node can collect the physiological signals (e.g.: such as EEG, ECG, body temperature, blood pressure, heart beat etc.), when attached to a human body [3]. The processing unit processes all the sensed signals, then sends out the data based on communication protocols. All the processed data will be transmitted through a wireless link to a portable, personal base-station. Doctors can then obtain all the patients' data through the network.

The main challenge for such sensor node is the high power consumption of portable devices. A solution to this challenge is the integration of the portable devices' digital and RF circuitry into one chip.

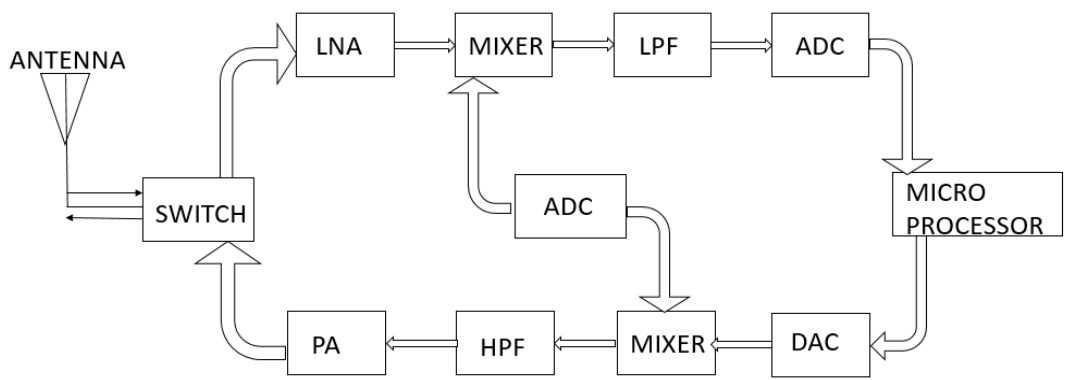

Figure 2. Block diagram of a transmitter

The receiver will receive the signal and will also perform DSP processing after the data is sent out by the transmitter [3]. Figure 2 is the transmitter diagram. It is desirable that the transmitter and receiver are low power devices. The director-conversion transmitter is very popular for such applications, because it offers versatility, flexibility, spectral efficiency, and low complexity. These features make the transmitter simpler than the super-heterodyne transmitter. Small chip and circuit size, and low power consumption can be achieved with a direct-conversion transmitter architecture. For the front-end transmitter, the major objectives are 1) transmit RF signals and 2) recover the biosignal classification. This paper proposes a low power receiver design. This paper is mainly for the power amplifier design, since other portions of the circuit design are already discussed in the paper [3]. In order to meet the standards, the PA is designed as shown in table 1.

Table 1: PA design requirement.

| Parameter | Target(Unit) |
|---|---|
| Gain | 10dBm |
| PAE | 10% |
| Stability | >1 |
| S11 | -10 dB |

## 2. METHODS

Over the past 30 years, research on CMOS radio-frequency (RF) front-end circuits has progressed extremely quickly. The ultimate goal for the wireless industry is to minimize the trade-offs between performance and cost, and between performance and low power consumption design [4].

The proposed Class AB amplifier has low output power and good linearity based on the IEEE 802.11b communication protocol. The class AB power amplifier topology is shown in figure 3. The 2.4GHz PA is a two stage common-source amplifier. The first stage is a driver stage, used for providing sufficient driving capability and a proper gain, as seen in figure 4(a), and the second stage is the power output stage which used for performing sufficient output power, as seen in figure 4(b)[5].

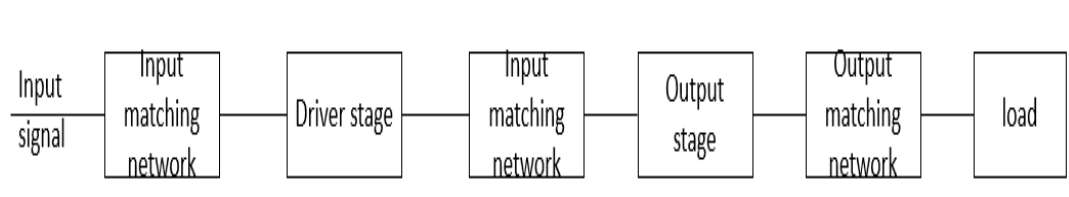

Figure 3. Block diagram of a class AB power amplifier

To ensure low cost, so the PA is designed via a CMOS process. And the initial equirements as seen in Table1.

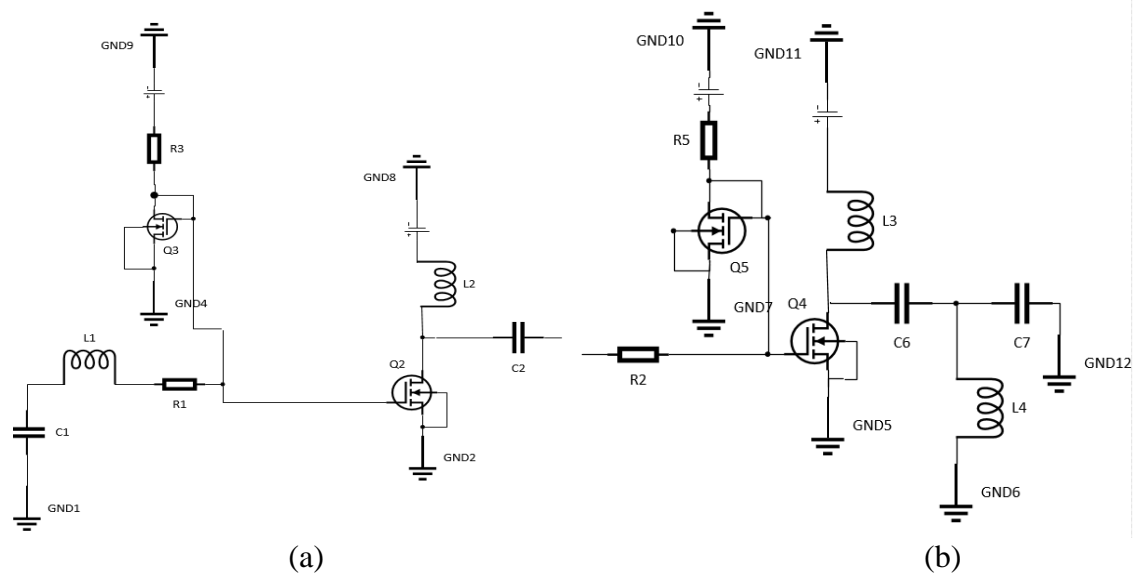

(a)                                                                                      (b)

Figure 4. (a) Schematic of driver stage (b) Schematic of output stage

For the drive-level circuit, the first design concern was ensuring the input and output conjugate match to different sizes of CMOS transistor. To get the optimum bias, small-signal simulation and 1dB compression point simulation are completed by their power output capability. Resulting design values can be shown in Table 2 and Table3.

Table 2: 2.4GHz PA driver stage component.

| Parameter | Size (Unit) |
|---|---|
| Q2 | W/L=4.8um/1um (f=16,m=12) |
| Q3 | W/L=0.3um/1um |
| R1 | 14.5 Ohm |
| R3 | 13K Ohm |
| L1 | 22 nH(Q=20) |
| L2 | 15 nH (Q=20) |
| C1 | 200 fF |
| C2 | 10 pF |

After the output stage and driver stage, the inter-stage matching circuit is more challenging. If the input of second stage and output of the first stage are all conjugate matched to 50Ω, the two stages can be connected directly. The complete optimized circuit is shown in Figure 5.

Table 3: 2.4GHz PA Output Stage Component.

| Parameter | Size (Unit) |
|---|---|
| Q4 | W/L=4.8um/3um (f=16,m=12) |
| Q5 | W/L=0.3um/1.2um |
| R2 | 22.2 Ohm |
| R5 | 7K Ohm |
| L3 | 15 nH(Q=20) |
| L4 | 400 nH (Q=20) |
| C6 | 800 fF |
| C7 | 20 pF |

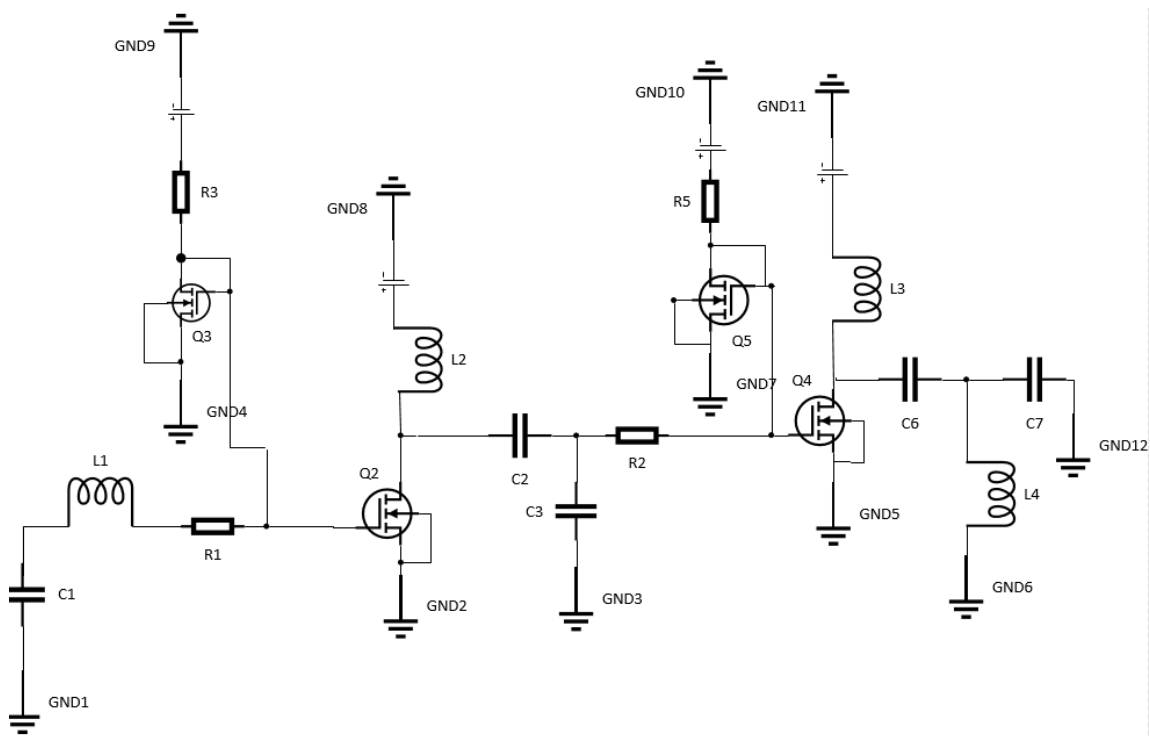

Figure 5. Overall circuit schematic

## 3. RESULTS

As seen in figure 6(a), the gain is 10. As seen in figure 6(b), the frequency is at 2.4 GHz the S11 is less than -10 dB, also, the total power of the PA is 0.135 W.

As seen in figure 7(a), Kf is larger than 1 for all frequencies from 1 to 3 GHz, so this circuit is totally stable. And the PAE is 7.5% at input power 0 dB.

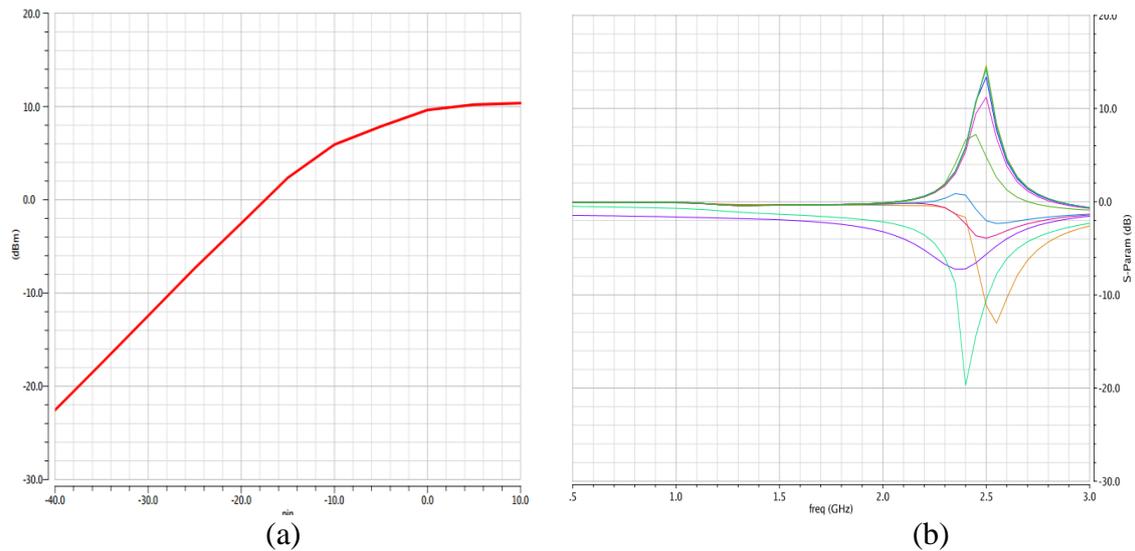

Figure 6. (a) Output power   (b) S11

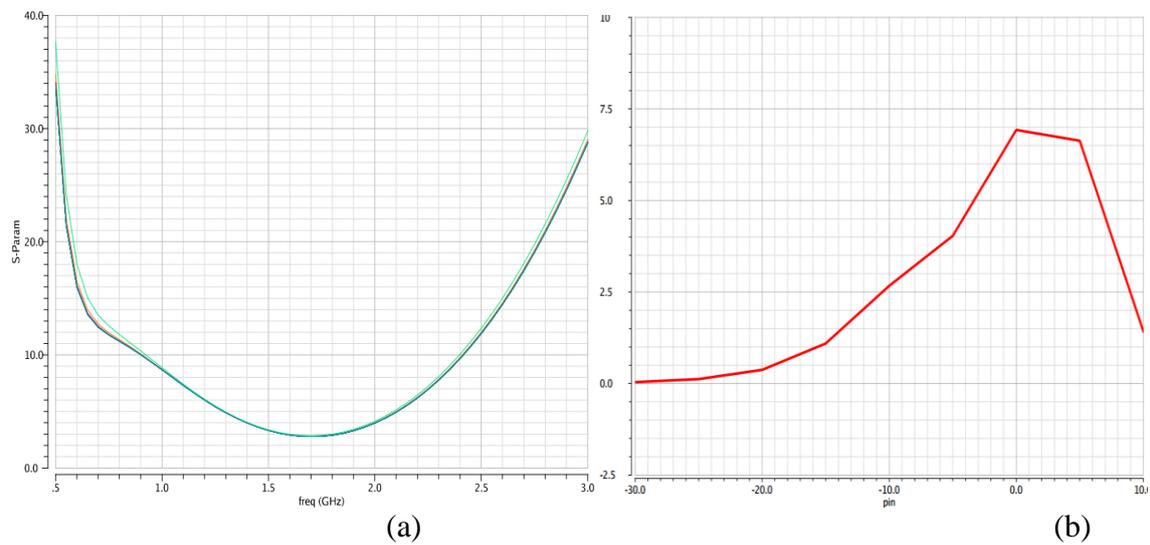

Figure 7. (a) Kf  (b)  PAE

## 4. CONCLUSION

This paper describes the method of designing and simulating power amplifier using cadence software based on SIMC CMOS process 180nm technology. This PA is used for sensor networks. This research is still in the early stages of development of a low cost and low power device. In order to reach the performance that is needed, the PA process uses group III and IV elements. This circuit meets the scheduled requirements for the CMOS process, but it still has room to improve performance metrics. When the sensor is coupled with communications technologies such as mobile phones and the Internet, the sensor network constant information flow between individuals and their doctors. Such low cost and low power device can save a lot of hospitalization resources. To realize this, future improvement is needed.

**AUTHORS**

Wei Cai is a graduate student at the University of California, Irvine, CA. She received her Masters degree from Dept. of Electrical Engineering, University of Hawaii at Manoa and Bachelor degree from ZhejiangUniversity, China. Her research interests include device physics simulation, analog/ RF circuit design.

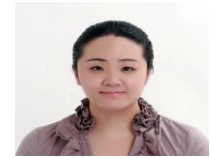

Liang huang is an associate Professor, Electronics College of Zhejiang Gongshang University. He got phd from Zhejiang University china, and finished his postdoc at Polytechnic of Turin, Italy, and Hanyang University, Seoul, Korea. His research is mainly focus on Research on Intelligent Control; Electrical Robotics.

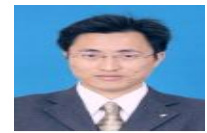

Wujie Wen is an assistant Professor at Department of Electrical And Computer Engineering of Florida International University. He got his Ph.D. in Electrical and Computer Engineering from University of Pittsburgh in 2015, his research is in the span emerging memory and next generation storage systems, VLSI circuit design and computer architecture, hardware acceleration (Neuromorphic computing) and hardware security

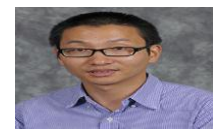